\documentclass[a4paper,twoside,fleqn]{article}
\usepackage{a4,epsfig}

\begin{document}
%\pagenumbering{arabic}
%\title{Thermal Stiffening of Curvature Elasticity in Fluid Membranes}
\title{Effect of Thermal Undulations on the Bending Elasticity and Spontaneous Curvature of Fluid Membranes}
\author{H.A.Pinnow\footnote{present address: Physics Department, University
    of Essen, D--45117 Essen, e-mail: pinnow@theo-phys.uni-essen.de},
  W. Helfrich\footnote{Wolfgang.Helfrich@physik.fu-berlin.de}
\\
{\em Fachbereich Physik, Freie Universit\"at Berlin\/}
\\
{\em Arnimallee 14, D--14195 Berlin, Germany\/}
\\
PACS.68.10.Et - Interface elasticity, viscosity and viscoelasticity
\\
PACS.82.65.Dp - Thermodynamics of surfaces and interfaces}

%\address{Freie Universit\"at Berlin
%Arnimallee 14, 14195 Berlin, Germany}

\maketitle

\begin{abstract}
We amplify previous arguments why mean curvature should be used as measure of
integration in calculating the effective bending rigidity of fluid membranes
subjected to a weak background curvature. The stiffening of the membrane by
its fluctuations, recently derived for spherical shapes, is recovered for
cylindrical curvature. Employing curvilinear coordinates, we then discuss
stiffening for arbitrary shapes, confirm that the elastic modulus of Gaussian
curvature is not renormalized in the presence of fluctuations, and show for
the first time that any spontaneous curvature also remains unchanged.
\\
\end{abstract}

%%% Local Variables: 
%%% mode: latex
%%% TeX-master: "art"
%%% End: 

\vspace{10mm}
%\begin{spacing}{2}
\section*{Introduction}
The fluid membranes of red blood cells and giant vesicles in water undergo microscopically visible
thermal shape fluctuations \cite{Bro}. The thermal undulations lead to
orientational decorrelation on a scale characterized by de Gennes' persistence
length \cite{DGT}. About fifteen years ago, one of us proposed that the
decrease of correlation should be accompanied by a softening of the
membrane. Specifically, the effective bending rigidity $\kappa^{\prime}$ was predicted
to obey  
\begin{equation}
  \label{0.1}
  \kappa^{\prime}\ =\ \kappa-\alpha \frac{kT}{8\pi}\mbox{log}M\ ,
\end{equation}
where $\kappa$ is the bare bending rigidity and $M$ the number of molecules
making up the membrane. The numerical prefactor was found to be $\alpha=1$ in
the first derivation \cite{hel2} and two later ones intended to support this
result \cite{hel3,hel4}. In all three papers mean curvature was assumed to be the
statistical measure. However, serious mistakes were made in the
calculations. Peliti and Leibler, whose result was the first to be published
\cite{P1}, and all other authors \cite{For1,K2,ami,Dav1,Dav2,BSK} rederived eq. (\ref{0.1}) for the
renormalized bending rigidity, but with the prefactor $\alpha=3$. They took normal
displacement of the membrane from its equilibrium shape as measure of integration.
Only recently the problem was reconsidered \cite{hel6} when the mistakes were
noted in the calculations employing mean curvature. In addition to correcting
the previous calculation for the sphere \cite{hel3}, a new
calculation was done for the almost flat fluctuating membrane. It is based on
a local bending mode expansion called hat model \cite{hel5}. In both cases it was
found that fluid membranes should be stiffened, not softened, by their thermal
undulations. The form of the relationship (\ref{0.1}) is maintained, while the
numerical prefactor changes sign, becoming $\alpha=-1$. However, it has be
mentioned that a few years ago Gompper and Kroll found a membrane softening in
Monte Carlo simulations of fluctuating vesicles \cite{GK}.\\
\\
In the present work we deal once more with the problem of selecting the
correct statistical measure, taking advantage of the analogies between polymers in
two-dimensional space and fluid membranes. Subsequently, we consider a particularly transparent
situation, cylindrical curvature, to rederive the effective bending rigidity
in this case. Employing curvilinear coordinates, we then find the
stiffening, i.e. the validity of eq. (\ref{0.1}) with $\alpha=-1$, for more general
deformations of the initially flat membrane. In the same
tensor notation, we confirm that the elastic modulus of Gaussian
curvature is not affected by thermal undulations and show for the first
time that this also holds for any spontaneous curvature of the membrane.
\\

%%% Local Variables: 
%%% mode: latex
%%% TeX-master: "art"
%%% End: 

\section*{Formulation of the problem}
The bending energy of a fluid membrane may be expressed by the Hamiltonian
\begin{eqnarray}
  \label{1.0.0}
  H=\int \mbox{d}^2\sigma \ \sqrt{g} \left(\frac{1}{2} \kappa J^2-\kappa c_s J+ \bar{\kappa}K \right)\ ,
\end{eqnarray}
where $\kappa$ and $\bar \kappa$ are the bending moduli and $c_s$ is the spontaneous curvature.
The terms in parentheses represent the surface energy density and $\sqrt{g}\
\mbox{d}^2\sigma$ is the local surface element of the membrane. Let $\vec
X(\bar \sigma)\ $, $\bar \sigma=(\sigma^{1},\sigma^{2})$, be a parametrization
of the membrane surface with the contravariant coordinates $\sigma^{1}$ and $\sigma^{2}$. Then $\mbox{d}^2 \sigma=\mbox{d}\sigma^{1}\mbox{d}\sigma^{2}$ and $g=\mbox{det}\ g_{ij}\ $, where
\begin{displaymath}
  g_{ij}=\partial_i\vec X \cdot \partial_j\vec X\ ,\quad \textrm{with} \quad
  \partial_i\vec X=\partial \vec X / \partial \sigma^i \ ,\quad i,j=1,2\ ,
\end{displaymath}
are the coefficients of the first fundamental form.
$J$ and $K$ denote trace and determinant of the extrinsic curvature tensor $J^i_j= g^{ik}J_{kj}$, respectively, where
\begin{displaymath}
  J_{ij}=\vec N \cdot \partial_{ij} \vec X \ ,\quad i,j=1,2\ 
\end{displaymath}
are the coefficients of the second fundamental form, $\vec N$ being (in
the case of a closed surface) the 
outward unit normal vector to the surface. In other words,
\begin{displaymath}
  J= c_1+c_2 \quad \textrm{and} \quad K=c_1 \cdot c_2\ ,
\end{displaymath}
are twice the mean curvature and the Gaussian curvature, respectively, with
$c_1$ and $c_2$ being the principal curvatures. Note that with this convention
$J$ is negative for spheres and cylinders.
\\
\\
The concept of an effective rigidity can be understood as follows: Starting
from a basically flat, weakly fluctuating piece of membrane, we are interested
in the free energy $\Gamma$ needed to deform the piece into a fluctuating
shape of nonvanishing background curvature $J$. It turns out that the effect
of the short wavelength fluctuations upon the elastic behaviour on large
scales can be absorbed into the material parameters of the bending Hamiltonian
(\ref{1.0.0}). Accordingly, the contribution to $\Gamma$ from the first term
in (\ref{1.0.0}) may be written as
\begin{equation}
  \label{1.0.1}
  \Gamma\ =\ \frac{1}{2}\kappa^{\prime} \int J^2 \sqrt{g}\ \mbox{d}^2\sigma\ .
\end{equation}
Here the curvature and the surface element refer to the nonfluctuating
background and $\kappa^{\prime} \ne \kappa$ is the effective bending
rigidity.\\
As is common in shape calculations, the membrane is assumed to be
unstretchable. The free energy $\Gamma$ refers to a background surface of fixed
area. The associated real membrane area is slightly larger because of the
bending fluctuations. Any shrinkage of base area resulting from the
fluctuations can be formally avoided by introducing a (fictitious) reservoir
of basically flat membrane area. Alternatively, $\Gamma$ can be taken to refer
to the shrunken base area. Such a shrinkage does not affect the bending
energy of the base if it is done by an isotropic scale transformation, as
happens automatically in the case of a freely fluctuating spherical
vesicle. Fortunately, these subtleties seem to be irrelevant in practice. 
\\
The free energy of bending, $\Gamma$, can be obtained in two different, but
equivalent ways. One may "freeze"  the undulations in an instantaneous
configuration and then calculate the bending energy of the rippled membrane at
fixed undulation amplitudes. The squared fluctuation mode amplitudes are
averaged only afterwards.\\
Alternatively, one may start from the free energy of the fluctuating flat
membrane and investigate how it changes when the membrane is subjected to a
background curvature $J \ne 0$. The fluctuation dependent part of the bending energy $\Gamma$ is now entropic,
resulting from the effect of $J$ on the mean square amplitudes of the
undulation modes.\\
In both cases we are faced with the problem of choosing the right measure of
integration. The undulations as expressed by the measure have to be kept constant
when the membrane with "frozen" undulations is deformed in the
large. Correspondingly, the entropic part of the free energy of background
deformations has to be calculated from the change  of the mode mean square
amplitudes in terms of the measure. We will expand now on some of the
arguments why mean curvature, or any quantity proportional to it, is
a correct measure of integration. The scale of the measure is of no interest
because we are dealing with differences, not absolute values of entropy.

\section*{Considerations on the measure of integration}
A heuristic argument for choosing mean curvature as statistical measure is the
fact that it can be regarded as the physical strain of the fluid
membrane, much like curvature in the case of a (stiff)
polymer. Accordingly, the superposition of two deformations, e.g. uniform background curvature
and a (sinusoidal) ripple, begins with adding the strains as they are in the
absence of the other deformation. The simultaneous presence of both
deformations will produce coupling terms in the effective total bending energy that are
quadratic in both the background curvature and the ripple amplitude. Employing
the strain as statistical measure has the particular attraction of avoiding
spurious coupling terms. They arise from a change of the ripple amplitude in
terms of the strain when other measures are used such as normal displacement
from the background or reorientation of the layer normal.\\
\\
From a technical point of view, mean curvature can serve as measure of
integration only if its local values determine the shape of the polymer or
membrane, respectively. A one--to--one relationship between shape and local
curvatures (which may be molecular) clearly exists for the polymer in the
plane, a simplified model system frequently considered in the following. For
membranes, a similar invertible mapping is particularly transparent in the hat
model which expands the fluctuations of the infinite planar membrane into
local mean curvature modes\cite{hel6}. There are only minor restrictions to the local
mean curvature fluctuations in other geometries, such as periodic boundary
conditions, cylinders or spheres (see below). However, any (small) membrane
deformation compatible with the boundary condition can be completely specified
either by local normal displacements or mean curvatures. In all these cases
Gaussian curvature fluctuates together with mean curvature, but the former is
completely slaved by the latter because of the boundary conditions. The
general possibility of invertible mappings between shape and mean curvature,
including local area conservation and lateral flow, has been discussed
elsewhere \cite{hel4,hel6}.\\
\\
Finally, there is a cogent reason to select (mean) curvature as
statistical measure for polymers in $2$D or fluid membranes. It
stems from the fact that in both systems every local strain fluctuates
independently of all the others (apart from the minor restrictions just
mentioned). As a result, the total fluctuation entropy is the sum of the equal
entropies of all the molecular (or other local) fluctuations. Independence and
equality imply that the entropy $S_{q}$ of a sinusoidal (mean) curvature mode
on a flat background does not depend on its wave vector $q$. This is because for all $q \neq 0$
the molecules are in the same way homogeneously distributed over the phases of the sine
wave. The $q$--independence of $S_{q}$ indicates that mean curvature is a
suitable measure of integration since the mean square curvature amplitudes
also do not depend on $q$. Let us add as a remark that the absence of a wave
vector dependent Jacobian can perhaps be regarded as characteristic of a
correct measure.
\\
The condition of the mode entropy being independent of $q$ is violated if the
entropy is calculated in terms of other statistical measures. Taking normal
displacement, one generally derives
\begin{displaymath}
  S_{q}=\frac{1}{2}k \log q^{4}\ + \ const.\ ,
\end{displaymath}
from the variation with $1/q^{4}$ of the mean square displacement amplitude of a sinusoidal fluctuation mode. The $q$--dependent part of $S_{q}$ is half as strong if
normal displacement is replaced by the tilt angle of of the polymer or
membrane with respect to the background. Evidently, neither normal
displacement nor tilt angle is a correct measure of integration. It should
be noted that in many statistical--mechanical calculations the choice of the
measure is irrelevant. However, it matters whenever the coupling of two
bending deformations is considered as will be demonstrated below.
\\
After these general statements let us address particular model systems to
deal with the statistical mechanics of both polymers in a plane and fluid
membranes. They will confirm the claim that mean curvature is a correct
measure of integration for fluid membranes. Examples of the invertible mapping between membrane
shape and local mean curvature will also be given, including the restrictions
on the latter.\\
\\
We first have a look at the comparatively simple problem of a polymer chain
fluctuating in a plane \cite{freed}. The configurational space of the polymer is
swept completely by varying the angles between adjacent
links. Disregarding fluctuations of link length, one may speak of local
curvatures instead of angles. Each joint between subsequent links has its own
free energy arising from its curvature fluctuations. Although the change of a
particular angle rotates the two sides of the polymer chain with respect to
each other, it does not affect their free energies. We assume here that the
polymer is almost straight. In a continuum
description such a polymer is equivalent to a (thin) rod or wire.\\
Using local curvature as the measure of integration in dealing with
fluctuating polymer chains is not in conflict with the fundamental measure
which for mass points is translation. Local curvature is a shorthand notation
for relative displacement if, e.g., each joint of the polymer chain is
regarded as a mass point. Then a (small) normal displacement of point $n+1$
from the tangent to the polymer defined by the two preceding points is
proportional to the curvature at point $n$. In contrast, the absolute normal
displacement of a nearly straight fluctuating polymer from a base line is not
a suitable statistical measure. It implies a freedom of motion for each
individual mass
point which actually is suppressed by cohesion and bending stiffness. The
remaining freedom is fully accounted for by the curvature.
\\
The polymer allows a simple demonstration of the effect of a wrong statistical
measure on the curvature amplitude of a ripple. We consider a sinusoidal wave
on a polymer with a uniform background curvature $-1/r$, the negative sign
indicating a downward bend. A sketch of the situation is given in Fig. 1. Let
us keep all normal displacements $\nu(s)$ fixed, $s$ being the arc length of
the background, while we bend the polymer as a whole. This amounts to using
the displacement measure. The procedure results in a weakening of the ripple
curvature because the total curvature changes from $\mbox{d}^{2}\nu
/\mbox{d}s^{2}$ to $-1/r+\mbox{d}^{2}\nu
/\mbox{d}s^{2}+\nu/r^{2}$ up to linear order in $\nu$, with the two terms
constituting the ripple curvature being of opposite sign. Clearly, the change
of the ripple curvature amplitude is physically not acceptable so that normal
displacement does not appear to be the correct statistical measure.
\\
%Let us finally point out that by choosing the link angles we actually
%parametrize the configurational space of the polymer chain by canonic
%variables, which automatically respect the restriction of fixed link lengths. 
%Liouville's theorem then guarantees that in the functional integral
%the fundamental measure $\mathcal{D}[p]\mathcal{D}[q]$ is correct. That is, a
%priori to each volume element in the space $(p,q)$ is attibuted the same
%statistical weight. Integrating out the conjugate momenta only contributes a
%factor to the free energy that can be absorbed into the definition of the
%remaining measure $\mathcal{D}[q]$. On the other hand, starting from the basic
%translational measure $\mathcal{D}[\nu]$ necessitates inference of a measure factor due to the
%projection of volume elements $\mathcal{D}[\nu]$ onto allowed configurations being
%parametrized by the curvature $\xi$. The corresponding Jacobian reads in our notation
%$|\partial \xi / \partial \nu |$ (compare with eq. \ref{1.2}).
\\
Continuing the two--dimensional polymer into the third space dimension results
in a bent membrane with zero curvature in the new direction. The similarity of
such a membrane with the polymer in a plane strongly suggests that some kind
of curvature rather than displacement from a fixed base should be the right
measure of integration for fluid membranes.
\\
In order to extend the preceding argument to membranes that can be
bent not only in a single direction let us consider a fluid membrane made of mass
points. A particularly simple situation arises for the almost flat, weakly
fluctuating membrane: Like the polymer, the idealized fluid membrane can be
build up by adding the constituent mass points one by one. For easy
statistics, we may think of a (local) lattice of mass points, quadratic or
hexagonal, which is formed row after row. A new molecule continuing a straight
sequence completes a plaquette comprising a certain number of  preexisting
neighbours. If the number is large enough the plaquette can be used to define
the mean curvature at its center. Five molecules is the minimum; they may be
thought to form a centered square, as is illustrated in Fig. 2. The height of
the last molecule added, relative to a suitably chosen tangent plane, finalizes the value of
the curvature $J$. On a quadratic lattice, $J$ may be taken (in lowest order)
to be the sum of the (normal) curvatures of the two diagonals of the
plaquette. Let us emphasize, that it
is not the relative height of the final molecule that gives a parametrization
of the phase space of the particle in the middle of the plaquette,
but the simultaneously varying mean curvature. This quantity is invariant with respect to the
direction along which rows are formed one by one. The tangent plane may be replaced by the base plane,
shifted to the local height, if the fluctuations are weak enough to leave the
membrane almost flat. Obviously, the last simplification is possible in the
example of Fig. 2.
\\
Regardless of the measure of integration, it is convenient to describe the
thermal undulations superimposed on a background shape of given $J$ and $K$ in terms of normal
displacement $\nu(\bar \sigma)$. To linear order, $\nu$ and the associated
extra curvature $\xi=J^{\prime}-J$ are related through
\begin{equation}
  \label{1.2}
  \xi=[J^2-2K+D^2]\ \nu\ ,
\end{equation}
$D^2$ being the Laplace--Beltrami operator (see eq. (\ref{2.4}) below and
subsequent definitions). The difference $\xi$ is equivalent to $J$ as
statistical measure.\\
Undulation modes can be defined as eigenfunctions of the operator represented
by the bracketed expression. For surfaces of uniform background curvature,
such as planes, cylinders and spheres, this ensures that for each mode the
functions $\xi(\bar \sigma)$ and $\nu(\bar \sigma)$ are proportional to each
other in a first order approximation, the factor of proportionality depending
on the wave number. The transformation (\ref{1.2}) can be inverted whenever
for a given undulation mode the operator in brackets does not
vanish or affect the membrane area linearly in the mode amplitude. Well--known examples of such exceptions are the three $l=1$ modes of
the fluctuating sphere which actually represent translations and the $l=0$
mode which requires area changes of the unstretchable membrane.
With periodic boundary conditions the only lost mode is uniform mean curvature
($q=0$) as it is incompatible with the boundary conditions. In the case of the
cylinder two long--wavelength curvature modes are lost because they represent
translations normal to the cylinder axis. Periodic boundary conditions along
the axis and area conservation suppress two other mean curvature modes. The
loss of modes implies that the local mean curvatures are not totally free to
fluctuate independently of each other. However, these restrictions seem
irrelevant in the present context for membranes comprising large
numbers of molecules.
\\

%%% Local Variables: 
%%% mode: latex
%%% TeX-master: "art"
%%% End: 

\section*{Basic formulas}
In order to calculate the bending Hamiltonian (\ref{1.0.0}) for a
fluctuating membrane, we first present expressions for the mean curvatures and
the surface elements of membrane shapes lying close to a given background
shape. The latter is also referred to as the projected shape.   
\\
If $\vec X(\bar \sigma)\ , \bar \sigma=(\sigma^{1},\sigma^{2})$, is a
parametrization of the background configuration, any slightly deviating
configuration $\vec X^{\prime}(\bar \sigma)$ may be represented in normal gauge \cite{For1,Dav1,K3} by
\begin{equation}
  \label{2.2}
  \vec X^{\prime}(\bar \sigma)=\vec X(\bar \sigma)+\nu(\bar \sigma)\vec N(\bar \sigma)\ ,
\end{equation}
$\vec N(\bar \sigma)$ being the local outward unit normal vector to
$\vec X(\bar \sigma)$. From (\ref{2.2}) we find that up to quadratic order in
$\nu$ the surface element of the deformed membrane obeys (cf.\cite{For1,K2,K3,Ou,Diff3})
\begin{equation}
  \label{2.3}
  \sqrt{g^{\prime}}=\sqrt{g}\,(1-\nu\,J+\frac{1}{2}\,(D\,\nu)^2+K\,\nu^2)\ .
\end{equation}
In the same approximation the curvature $J^{\prime}$ becomes
\begin{eqnarray}
  \label{2.4}
  \lefteqn{J^{\prime}=J+(D^2 +J^2-2K)\ \nu-\frac{1}{2}\ J\ (D \nu)^2} \nonumber \\
  &&+J^{ij}\ D_i \nu D_j \nu + (J^3-3JK)\ \nu^2+2J^{ij}\ \nu D_i D_j \nu+ \nonumber \\
  &&+\nu\ D_n \nu\ g^{ij} (D_j J^n_i+J\ \Gamma^n_{ij})\ ,
\end{eqnarray}
Here, $g_{ij}$ and $J_{ij}$ are the coefficients of the first and second
fundamental forms of the background surface, respectively. $\Gamma^l_{ij}$ are
the Gauss--Christoffel symbols of the second kind derived from
$g_{ij}$. $D_{i}$ is the covariant derivative %with respect to $g_{ij}$
being
defined as $D_{i}\nu=\partial_{i}\nu=\partial \nu/ \partial \sigma^{i}$, and
$D_{i}D_{j}\nu=\partial_{ij}\nu-\Gamma^{k}_{ij}\partial_{k}\nu$.
$D^2=g^{ij}D_iD_j$ is the Laplace--Beltrami operator and we use the abbreviation $(D\nu)^2=D^i\nu\ D_i\nu$. The
variation of the Gaussian curvature $K$ need not to be considered in (\ref{1.0.0}) thanks to the Gauss--Bonnet theorem \cite{Diff4}.

%%% Local Variables: 
%%% mode: latex
%%% TeX-master: "art"
%%% End: 

\section*{Renormalization of $\kappa$ for the cylinder via frozen mode amplitudes}

Let us now derive the effective bending rigidity for cylindrical background curvature in the
"frozen mode"  approach. 
We first write down reduced versions of (\ref{2.3}) and (\ref{2.4}) that apply
to this special case. Using geodesic coordinates $x=r\cdot \varphi$ and $z$ to
parametrize the cylinder, where $\varphi$ and $z$ are usual polar coordinates,
we find with $J=-1/r$: 

\begin{equation}
  \label{a.1}
  \sqrt{g^{\prime}}= 1+\frac{1}{r}\nu+\frac{1}{2}(\nabla \nu)^2 +O(\nu^{3}) \  
\end{equation}
and
\begin{eqnarray}\label{a.2}
  \lefteqn{J^{\prime}=-\frac{1}{r}+\frac{1}{r^2}\nu+\Delta \nu +\frac{1}{2}\frac{1}{r}(\nabla \nu)^2}\nonumber \\
  &&-\frac{1}{r}(\partial_x \nu)^2-2\frac{1}{r}\nu \partial_{xx} \nu-\frac{1}{r^3}\nu^2+O(\nu^{3})\ .
\end{eqnarray}
The linear part of $J^{\prime}$ in (\ref{a.2}) leads to the transformation
(\ref{1.2}) and its inversion for the cylinder,
\begin{equation}
  \label{a.3}
  (1/r^2+\Delta)\nu=\xi\ \Leftrightarrow \ \nu=\Delta^{-1}(1+1/r^2\Delta^{-1})^{-1}\xi\ ,
\end{equation}
where $\Delta$ is the usual Laplacian for a flat base. Obviously, the thermal undulations
on a cylinder can be expanded, like those on a flat background, into
sinusoidal waves which at the same time are curvature and normal displacement
modes.\\
We will consider a quadratic piece of membrane subjected to a weak
cylindrical background bend and satisfying periodic boundary
conditions. Alternatively, the
membrane may be closed to form a cylinder so that the edge of the square
equals the circumference in length. In the
displacement representation the sine and cosine deformation modes read
\begin{equation}
  \label{a.3.0.1}
  \nu=a_{\vec q}\sin(\vec q\ \vec r)\quad \textrm{and}\quad \nu=b_{\vec q}\cos(\vec q\ \vec r)\ ,
\end{equation}
where $\vec r=(x,z)$ and $\vec q=(q_{x},q_{z})$ with $q_{x}=2\pi n/L$ and
$q_{z}=2\pi m/L$. The numbers $n$ and $m$ are running from $-L/(2a)$
to $L/(2a)$, but only over half the $\vec q$ plane. The constant mode
$(n,m)=(0,0)$ is excluded. $L$ denotes the size of the background shape and
$a$ is the molecular distance in the membrane. 
The operator acting on $\nu$ in (\ref{a.3}) is $(1/r^{2}-\vec q^{\ 2})$ for a given mode.\\
In order to renormalize the bending rigidity let us first consider the effect
of a single sinusoidal mode. 
A straightforward calculation starting from (\ref{a.1}) and (\ref{a.2}) and
involving a partial integration then
results in the Hamiltonian
\begin{eqnarray}
  \label{a.3.2}
  \lefteqn{\int \frac{1}{2}\kappa
    J^{\prime 2}\sqrt{g^{\prime}}\mbox{d}x\mbox{d}z=}\nonumber\\
  &&\frac{1}{2}\kappa
  \left[\frac{1}{r^{2}}+\frac{1}{2}(1/r^{2}-\vec q^{\ 2})^{2}a_{\vec
      q}^{2}+\frac{1}{2}\frac{1}{r^{2}}(-\frac{1}{2}q_{x}^{2}+\frac{3}{2}q_{z}^{2}-\frac{1}{r^{2}}) a_{\vec q}^{2}\right]A_{base}\nonumber\\
  &&\ 
\end{eqnarray}
where $A_{base}=L^{2}$ denotes the area of the nonfluctuating
background shape.
Terms linear in $\nu$ do not occur as they vanish on integration. Of
the three terms in the brackets, the second one equals the mean value of
$\xi^{2}$, thus giving the
regular bending energy of the mode. Note that if the curvature measure
applies, the curvature difference $\xi$ is kept constant in the "frozen mode"
approach when the membrane is deformed. Accordingly, the third term is due to the coupling
of the mode curvatures to cylindrical curvature.\\
The statistical averages of the squared mode amplitudes are obtained from the equipartition theorem,
\begin{equation}
  \label{a.3.3}
\left< a_{\vec q}^{2}\right> =\left< b_{\vec q}^{2}\right> =\frac{2\ kT}{A_{base}\kappa(1/r^{2}-\vec q^{\ 2})^{2}}  \ .
\end{equation}
Inserting (\ref{a.3.3}) in (\ref{a.3.2}) and absorbing the
third term of (\ref{a.3.2}) in the first leads to
\begin{equation}
  \label{a.3.4}
  \!\!\!\!\!\!\!\!\!\!\!\!\!\!\left< \int \frac{1}{2}\kappa
    J^{\prime
      2}\sqrt{g^{\prime}}\mbox{d}x\mbox{d}z \right>=\frac{1}{2}\kappa\left(1+\frac{-\frac{1}{2}q_{x}^{2}+\frac{3}{2}q_{z}^{2}-1/r^{2}}{(1/r^{2}-\vec q^{\ 2})^{2}}\frac{kT}{A_{base}\kappa}\right)\cdot \frac{1}{r^{2}} A_{base}+\frac{1}{2}kT \ .
\end{equation}
Adding up the contributions of all fluctuation modes to the renormalization of $\kappa$
yields to lowest order in $kT/\kappa$
\begin{equation}
  \label{a.3.5}
  \kappa^{\prime}\ =\ \kappa \left(1+\sum_{\vec
      q}\frac{-q_{x}^{2}+3q_{z}^{2}}{(\vec q^{\ 2})^{2}}\frac{kT}{A_{base}\kappa}\right)\ ,
\end{equation}
where $1/r^{2}$ is omitted because of $q>1/r$. The sum is assumed to be small as compared to
unity. It does not contain $(n,m)=(0,0)$ and some seemingly divergent low wave vector
modes. The upper limits
of $n$ and $m$ are large numbers. Replacing the sum by the double integral
\begin{displaymath}
  \sum_{\vec q}\ \longrightarrow \
  \frac{A_{base}}{8\pi^{2}}\int^{\pi/a}_{2\pi/L}\mbox{d}^{2}\vec q\ ,
\end{displaymath}
we arrive with good
accuracy at
\begin{equation}
  \label{a.3.6}
  \kappa^{\prime}\ =\ \kappa +\frac{kT}{8\pi}\log M\ ,
\end{equation}
$M$ being the number of molecules in the membrane (or half this number
  for the bilayer). Evidently, we have
recovered the result derived previously for the fluctuating spherical vesicle,
i.e. eq. (\ref{0.1}) with $\alpha=-1$.
\\
The effect of thermal undulations on the bending rigidity of fluid membranes
was attributed to a coupling between the mode curvatures and the background
curvature. To confirm the role of coupling, we may for a moment compensate the uniform
background curvature by a spontaneous curvature. For the cylinder this means
$c_{s}=-1/r$ and $J^{\prime}-c_{s}=\xi$. Accordingly, the surface energy
density is simply
$\frac{1}{2}\kappa\xi^{2}=\frac{1}{2}\kappa\left((\frac{1}{r^{2}}+\Delta\right)\nu)^{2}$
up to quadratic order in $\xi$. The additional quadratic terms arising in the
case of interest, $c_{s}=0$, can therefore be regarded as coupling terms
that give rise to a renormalization of the bending rigidity. 
\\
For a direct understanding of the coupling let us consider two special types
of ripples, those exactly parallel or orthogonal to the cylindrical background
curvature.\\
{\em Parallel ripple:\/}\ \ \ \ When the ripple is parallel we can use the
analogy to a polymer in a plane. This makes it attractive to use a more direct
approach than the general calculation based on $\nu(x,y)$. Starting from a straight and stiff polymer or
a wire,
we impose on it a periodic deformation
\begin{displaymath}
  \xi(s)\ =\ c_{q}\ \sin(q\ s)\ ,
\end{displaymath}
where $\xi$ is curvature, $s$ arc length and $q$ wave vector. The undulation
diminishes the projected or background length of the wire. Alternatively, it
increases the total arc length $L^{\prime}$ of the wire (when connected to a reservoir) at fixed
projected length $L$. In both cases the ratio of the two lengths is
\begin{equation}
  \label{a.3.7}
  L^{\prime}/L\ =\ 1+\frac{1}{4}\frac{\langle  c_{q}^{2}\rangle}{q^{2}}
\end{equation}
to lowest order in $\xi$.\\
A frozen bending mode of the polymer is similar to a wire with a permanent
wave. Any sinusoidal bending deformation allows a given angle difference
between the ends of the wire to be spread over $L^{\prime}$ instead of
$L$. This reduces the associated bending energy density by $(L/L^{\prime})^{2}$
and the total bending energy by $L/L^{\prime}$ as compared to a polymer or
wire without waviness. The analogous ratio for parallel ripples in membranes
is $A/A^{\prime}$, where $A^{\prime}$ and $A$ are the real and projected
areas, respectively. Absorbing the effect into a renormalized bending
rigidity, we have $\kappa^{\prime}=\kappa\ A/A^{\prime}$. This agrees exactly
with what we obtain from (\ref{a.3.4}) for a single mode of type $(q_{x},0)$ if we
make use of $A^{\prime}=A(1+\frac{1}{4}q_{x}^{2}\ a_{q_{x},0}^{2})$. The
negative sign in front of the correction term in $\kappa^{\prime}$ indicates the
softening in this case.\\
An amusing illustration of the effect of parallel undulations is a rather
stiff polymer closed to form a ring. Its thermal undulations in terms of
curvature are not affected by the uniform bend and vice versa. Only the
decrease of the ring radius in the presence of undulations signals a lowering of the effective bending
stiffness of the polymer.
\\
{\em Orthogonal ripple:\/}\ \ \ \ Ripples orthogonal to the cylindrical base
curvature stiffen the membrane, in contrast to parallel ones. This is due to
the fact that at the crests of such a ripple the extra curvature $\xi$ is of
the same sign as the cylindrical curvature, and at the same time the membrane
area is increased as compared to the background shape.
In linear approximations the two effects obey
\begin{displaymath}
  J^{\prime 2}=\frac{1}{r^2}-2\frac{1}{r}\left(\frac{1}{r^2}+\partial_{zz}\right)\nu\ +\ O(\nu^2)
\end{displaymath}
and
\begin{displaymath}
  \sqrt{g^{\prime}}=1+\frac{\nu}{r}\ + \ O(\nu^2)\ .
\end{displaymath}
Calculating from these two expressions the total bending energy of a membrane
with a single excited mode of wave vector $(0,q_{z})$, we find the renormalized bending
rigidity of the cylinder to be
\begin{equation}\label{a.3.8}
  \kappa^{\prime
    }=\kappa(1-(-q_{z}^{2})\ a_{0,q_{z}}^{2}) .
\end{equation}
According to this formula the orthogonal ripple stiffens the membrane. It is
four times as effective as a softening, parallel ripple of equal wavelength.
\\
There is a second correction quadratic in $\nu$ to the bending energy due to the coupling between background curvature
and orthogonal ripple. It originates from the sloped regions of the ripple where
the membrane partially escapes  cylindrical curvature. The reduced principal
curvature is $(\cos\varphi) /r=(1/r)(1-\varphi^{2}/2+O(\varphi^{4}))$, $\varphi$ being the slope
angle. For a single mode of type $(0,q_{z})$ the second effect and the increase of
membrane area due to the slope result in a softening given by
\begin{equation}
  \label{a.3.9}
  \kappa^{\prime}=\kappa(1-\frac{1}{4}q_{z}^{2}\ a_{0,q_{z}}^{2})\ .
\end{equation}
Combining the opposite rigidity corrections of (\ref{a.3.8}) and (\ref{a.3.9}) results
in a stiffening which agrees with (\ref{a.3.4}).
\\
At this point let us briefly consider what would be different in a derivation
of $\kappa^{\prime}$ based on the displacement measure. Keeping $\nu$ fixed
instead of $\xi$ while the background is cylindrically bent implies that the
regular part of the bending energy density is $1/2\ \kappa(\Delta \nu)^{2}$
instead of $1/2\ \kappa(\Delta \nu-\nu/r^{2})^{2}$. Accordingly, the single
mode Hamiltonian (\ref{a.3.4}) would have to be recast in the form
\begin{eqnarray*}
  \lefteqn{\int \frac{1}{2}\kappa
    J^{\prime 2}\sqrt{g^{\prime}}\mbox{d}x\mbox{d}z=}\\
  &&\int\frac{1}{2}\kappa
  \left(\frac{1}{r^{2}}+(\vec q^{\ 2})^{2}\nu^{2}+\frac{1}{r^{2}}(-\frac{5}{2}q_{x}^{2}-\frac{1}{2}q_{z}^{2}+O(r^{-2}))\nu^{2}\right)\mbox{d}x\mbox{d}z \ , 
\end{eqnarray*}
if again the third term is to represent the coupling of the undulations to
cylindrical background curvature. The negative signs before $q_{x}^{2}$ and
$q_{z}^{2}$ in the coupling term suggest that both parallel and perpendicular ripples soften the
membrane. Absorbing the third term into the first renormalizes $\kappa$. The
same manipulations as above, including thermal averaging and summation over all modes, would lead
to
\begin{displaymath}
  \kappa^{\prime}\ =\ \kappa-3\frac{kT}{8\pi}\log M\ .
\end{displaymath}
This is the result obtained by all other authors. They generally used the
displacement measure and took the approach via undulation mode entropies.
\\
The undulations of unstretchable fluid membranes must be
accompanied by lateral displacements. They can be analyzed mode by mode and
consist of local and global parts. The local part redistributes membrane
material between the crests and troughs of a ripple in a bent surface and in
general varies linearly with the mode amplitude. The global part takes account
of the area absorbed by the undulation and goes with the square of the
amplitude. We have disregarded lateral motion, with one exception: In the case
of the wavy polymer it is automatically included by employing arc length as
independent variable. Inspection shows that lateral displacement need not be
taken into account in calculations of the effective bending rigidity of polymers and
membranes in the usual approximation that is quadratic in the mode amplitudes
.

%%% Local Variables: 
%%% mode: latex
%%% TeX-master: "art"
%%% End: 

\section*{Renormalization of $\kappa$, $c_{s}$ and $\bar \kappa$ via undulation mode entropies}
The effective bending rigidity has been derived above for uniform cylindrical
curvature and earlier for the spherical membrane, both times in the "frozen
mode" approach. The same stiffening was obtained in both calculations. Let us
try now to extend the result to more general geometries, showing
simultaneously that the undulations do not affect the spontaneous curvature
$c_{s}$ and the modulus of Gaussian curvature, $\bar \kappa$, of the
membrane. Obviously, we have to limit ourselves to "compact" shapes such as
squares, circles and spheres. For the following calculations it is
advantageous to follow the approach based on fluctuation mode entropies. The
curvature increment $\xi=J^{\prime}-J$ will again be used as measure of
integration.\\
In order to calculate the free energy of the fluctuating membrane
and to renormalize its elastic parameters, we employ a field--theoretic
approach as has been introduced by F\"orster \cite{For1} in this
frame. Considering some background configuration with parametrization $\vec
X(\bar \sigma)$, which describes the effective shape, we expand the bending Hamiltonian in (\ref{1.0.0}) to second order
around $\vec X(\bar \sigma)$. We drop linear terms, assuming the
background to be an equilibrium shape. The free energy $F$ of $\vec X$ may
then be written as \cite{For1,Dav1,K4}
\begin{equation}
  \label{0.6}
  F(\vec X)=H(\vec X)+\frac{1}{2}kT\log \det \frac{\delta^2 H(\xi)}{\delta \xi(\sigma)\delta \xi(\sigma^{\prime})}+O((kT)^2)\ ,
\end{equation}
where $\xi$ parametrizes the configurations $\vec X^{\prime}$ in the vicinity of
$\vec X$. After expanding the r.h.s. in powers of background curvature one may
introduce effective elastic parameters marked by a prime, such as
$\kappa^{\prime}$, and an effective Hamiltonian $H^{\prime}$ employing the new
parameters in conjunction with $J$ and $K$ of the background shape $\vec X$.
\\
To perform this program we first have to calculate the bending energies of the
fluctuating membrane. Starting from eq. (\ref{2.4}) we obtain
\begin{eqnarray}
  \label{4.1}
  \lefteqn{J^{\prime 2}=J^2+2J(D^2 +J^2-2K) \nu+((D^2 +J^2-2K) \nu)^2} \nonumber \\
  &&-J^2\ (D \nu)^2+2JJ^{ij}\ D_i \nu D_j \nu +2J(J^3-3JK)\ \nu^2 \nonumber \\
  &&+4JJ^{ij}\ \nu D_i D_j \nu+2J\nu\ D_n \nu\ g^{ij} (D_j J^n_i+J\ \Gamma^n_{ij})+O(\nu^{3})\ .
\end{eqnarray}
Combining this with (\ref{2.3}) yields
\begin{eqnarray}
  \label{4.2}
  \lefteqn{J^{\prime 2}\sqrt{g^{\prime}}=[J^2-J^3\nu+2J(D^{2}+J^2-2K)\nu-\frac{1}{2}J^2(D\nu)^2}\nonumber\\
  &&-2J^2\nu(D^{2}+J^2-2K)\nu\nonumber\\
    &&+4JJ^{ij}\nu D_i D_j\nu+2JJ^{ij}D_i\nu D_j\nu+((D^{2}+J^2-2K)\nu)^2\nonumber\\
    &&+JK\nu^2+2J(J^3-3JK)\nu^2+2J\nu D_n\nu
    g^{ij}(D_jJ_i^n+J\Gamma_{ij}^n)+O(\nu^{3}) ]\sqrt{g}\ .\nonumber\\
&&\
\end{eqnarray}
We also need
\begin{eqnarray}
  \label{4.3}
  \lefteqn{J^{\prime}\sqrt{g^{\prime}}=[J+(D^{2}+J^2-2K)\nu-J^2\nu}\nonumber\\
  &&-J\nu(D^{2}+J^2-2K)\nu+2J^{ij}\nu D_iD_j\nu+J^{ij}D_i\nu D_j\nu\nonumber\\
  &&+JK\nu^2+(J^3-3JK)\nu^2+\nu D_n\nu g^{ij}(D_jJ_i^n+J\Gamma_{ij}^n)+O(\nu^{3})]\sqrt{g}\ .
\end{eqnarray}
In the following the background curvatures $J$ and $K$ are assumed to be weak
and slowly varying. For all wave vectors except for the smallest ones the
predominant terms quadratic in $\nu$ are those that contain at least two
derivatives acting on $\nu$.
In calculating the effect of the short wavelength
fluctuations upon the long distance elastic behaviour of membranes comprising
a large number of molecules, it is sufficient to take into account only the
latter. Giving rise to cutoff dependent effects, they are the only ones that
can contribute to the renormalization of the elastic parameters. To replace the normal displacement $\nu$ by $\xi$ we
need the inversion of (\ref{1.2})

\begin{equation}
  \label{4.3.1}
  \nu=D^{-2}(1+(J^2-2K)D^{-2})^{-1}\xi\ ,
\end{equation}
In the spirit of the above simplification,
(\ref{4.3.1}) can be approximated for terms which are quadratic in $\nu$ at the stage of
$J^{\prime}$ and $g^{\prime 1/2}$ (see eqs. (\ref{2.3}) and (\ref{2.4})) by
\begin{displaymath}
  \nu=D^{-2}\xi \ .
\end{displaymath}
Disregarding irrelevant terms, we may thus write the bending energy as
%\pagebreak
\begin{eqnarray}
  \label{4.4}
  \lefteqn{\int\mbox{d}^2\sigma\ \sqrt{g^{\prime}} (\frac{\kappa}{2}J^{\prime 2}+c_s\kappa J^{\prime})=\int\mbox{d}^2\sigma\ \sqrt{g}[\frac{\kappa}{2}\cdot (J^2}\nonumber\\
  &&+2J\xi-J^3 D^{-2}(1+(J^2-2K)D^{-2})^{-1}\xi\nonumber\\
  &&+\xi(\frac{1}{2}J^2D^i D_iD^{-4}+2JJ^{ij}D_i D_j D^{-4} -2J^2 D^{-2}+I)\xi\nonumber\\
  &&+\kappa c_s\cdot (J+\xi+\xi (-JD^{-2}+J^{ij}D_iD_j /D^4)\xi)+ \textrm{irrel. terms}]\ , 
\end{eqnarray}
where we have performed some partial integrations, assuming either periodic
boundary conditions or undulations limited to a patch of membrane.
\\
We are now in a position to set up the functional integral sweeping curvature
space that gives the effective free energy $F$ of the background shape $\vec X$ with fields $J$ and $K$. In mechanical equilibrium linear terms in (\ref{4.4}) drop out
 so that we may write
\begin{eqnarray}
  \label{4.5.-2}
  \lefteqn{F=-kT\log\int {\mathcal D}[\xi]\exp\left[-\frac{1}{kT}\int
      \mbox{d}^2 \sigma \sqrt{g}\left(\frac{\kappa}{2}J^2-\kappa\ c_{s} J\right.\right.}\nonumber\\
  &&+\frac{\kappa}{2}\left(\xi\left(1+\frac{J^2}{2}D^{-2}+\frac{2JJ^{ij}D_iD_j}{D^4}\right.\right.\nonumber\\
        &&\left.\left.\left.\left.-\frac{2J^2}{D^2}+2c_s\frac{J^{ij}D_iD_j-JD^2}{D^4}+\ \textrm{irrel. terms}\right) \xi\right)\right) \right].
\end{eqnarray}

Extracting the background energy and integrating the multiple Gaussian we are led to \cite{K4} 
\begin{eqnarray}
  \label{4.5}
  \lefteqn{F=\int\mbox{d}^2\sigma\ \sqrt{g}\
    \left(\frac{\kappa}{2}J^2-\kappa\ c_{s} J\right)}\nonumber\\
  &&+\frac{kT}{2}\mbox{Tr}^{\prime}\log\left[\frac{\kappa}{2kT}\left(I+\frac{J^2}{2D^2}+\frac{2JJ^{ij}D_iD_j}{D^4}-\frac{2J^2}{D^2}+2c_s\frac{J^{ij}D_iD_j-JD^2}{D^4}\right)\right],\nonumber\\
  &&\ 
\end{eqnarray}
where $I$ is the identical operator. The prime on Tr indicates exclusion of some modes at the low wave number cutoff. We assume rather weak background bends, such that
\begin{equation}
  \label{4.5.0}
  J^2,K\ < \simeq \ 1/L^2\ ,
\end{equation}
and may thus expand the logarithm in powers of them. A further limitation of
$J^{2}$ may arise from $c_{s}$. The traces occuring in
(\ref{4.5}) are known to be \cite{Dav1,F}:
\begin{eqnarray}
  \mbox{Tr}^{\prime}\frac{J^2}{D^2}&\simeq& -\int\mbox{d}^2\sigma\ \sqrt{g}\ J^2\cdot \frac{2}{4\pi}\log\frac{L}{a}\qquad \textrm{and}\label{4.5.0.1}\\
  \mbox{Tr}^{\prime}\frac{2JJ^{ij}\ D_iD_j}{D^4}&\simeq& -\int\mbox{d}^2\sigma\ \sqrt{g}\ J^2\cdot\frac{2}{4\pi}\log\frac{L}{a}\label{4.5.0.2}\ ,
\end{eqnarray}
$L$ and $a$ being upper and lower cutoffs due to finite membrane size and
molecular distance, respectively. They are easily calculated for uniformly
bent shapes such as squares, spheres and cylinders. One obtains
%The dominant contributions in (\ref{4.5.0.1}) and (\ref{4.5.0.2}) is come from%the kernel in flat space. The free energy now reads
\begin{eqnarray}
  \label{4.6}
  \lefteqn{F=\int\mbox{d}^2\sigma\ \sqrt{g}\ \frac{1}{2}\kappa J^2\left(1+\frac{kT}{4\pi\kappa}\log\frac{L}{a}\right)+\frac{kT}{2}\mbox{Tr}^{\prime}\log\left(\frac{\kappa}{2kT}I\right)}\nonumber\\
  &&-\int\mbox{d}^2\sigma\ \sqrt{g}\ c_s\kappa J\cdot\left(1+\frac{kT}{4\pi\kappa}\log\frac{L}{a}\right)\ .
\end{eqnarray}
 The middle term of (\ref{4.6}) is equal to the free energy of
 the corresponding flat reference membrane. After subtracting it we are left
 with the free energy of bending which may be written in the renormalized form
\begin{equation}
  \label{4.7}
  \Gamma=\int\mbox{d}^2\sigma\ \sqrt{g}\
  (\frac{1}{2}\kappa^{\prime}J^2-\kappa^{\prime}c^{\prime}_sJ)\ ,
\end{equation}
where
\begin{eqnarray}
  \label{4.8}
  \kappa^{\prime}&=& \kappa +\frac{kT}{4\pi}\log\frac{L}{a}\ \\
  c^{\prime}_s&=&c_s\ .\label{4.9}
\end{eqnarray}
\\
For comparison with (\ref{0.1}), note that $L^2/a^2=M$.
Evidently, the effective bending rigidity comes out exactly as calculated for
the sphere \cite{hel6} and the cylinder, while the spontaneous curvature is not
modified by the thermal undulations.\\
Formula (\ref{4.7}) omits the Gaussian curvature
term of the bending energy. We dropped it in the unrenormalized bending energy since, as a
consequence of the Gauss--Bonnet theorem, the variation of
Gaussian curvature does not affect its integral if the membrane is closed or
satisfies periodic boundary conditions. Nevertheless, an additional Gaussian curvature
term could have emerged in the calculation of the effective bending rigidity,
as it does when normal displacement is the measure of integration
\cite{K3}. From the fact that such a term is absent in eq. (\ref{4.6}) we may infer that $\bar \kappa$ is not affected by
thermal undulations, i.e.
\begin{equation}
  \label{4.9.1}
  \bar \kappa^{\prime}\ =\ \bar \kappa\ ,
\end{equation}
in agreement with previous considerations \cite{hel6}.
%\\
%The above calculations could have been as well performed while staying always
%within the displacement representation. One then integrates over $\nu$ and
%takes care of curvature being the physical measure by infering a Jacobian
%factor $\det (\partial \xi (\bar \sigma)/ \partial \nu (\bar \sigma^{\prime}))$
%according to eq. (\ref{1.2}) into the functional integral.
\\
In the preceding calculation of the effective bending rigidity we did not specify
the shape of the membrane, except for stipulating that it be
"compact". Although the formula for $\kappa^{\prime}$ seems to be rather
robust, because of the logarithmic dependence on wave vector cutoffs, it is not expected to
hold for filamentous membranes. Specifically, the bending rigidity of the
membrane closed to form a cylinder will not stiffen indefinitely as the cylinder is
made longer. This is why we chose periodic boundary conditions with an axial
period as long as the circumference in our calculation for the cylinder.
\\
The renormalized elastic parameters calculated above are independent of the
base curvatures $J$ and $K$. It is an open question to which extent they
remain valid if these curvatures are nonuniform over the base area. In the
case of a wavy background one commonly uses as
upper limit in (\ref{4.6}) its wavelength or a fraction thereof instead of the
membrane size $L$. This is because ripples of larger wavelength are part of
the immutable background curvature.

%%% Local Variables: 
%%% mode: latex
%%% TeX-master: "art"
%%% End: 

\section*{Conclusion}
We have renormalized the bending elastic parameters of fluctuating fluid
membranes for cylindrical and more general membrane shapes. The same
logarithmic increase of the bending rigidity with membrane size was obtained
as previously for spherical vesicles. In addition, the modulus of Gaussian
curvature and, for the first time, spontaneous curvature were shown not to be
affected to lowest order by thermal undulations. Apart from the illustrative
treatment of the cylindrical membrane, we used arbitrary curvilinear
coordinates in the calculations to obtain generally valid results as far as
possible.
\\
The predicted stiffening of fluid membranes by their thermal undulations may
seem contrary to intuition, although there is no conflict with de Gennes'
\cite{DGT} derivation of the persistence length of membrane orientation. In
the case of polymers the loss of orientational correlation unavoidably results
in a breakdown of bending elasticity. This could be different in the case of
membranes because of their two-dimensional connectedness and, in particular,
the well-known scale invariance of bending energies. 
Moreover, it remains to be checked if the stiffening survives at very low bending
rigidities $(\kappa \approx kT)$, where the nearly flat approximation
is inapplicable at any scale.

%%% Local Variables: 
%%% mode: latex
%%% TeX-master: "art"
%%% End: 

\section*{Acknowledgement}
We are grateful for their interest to H. Kleinert, A.M.J. Schakel and
M.E.S. Borelli. W.H. thanks D. Nelson for comments and H.A.P. is indebted to
G. Foltin for enlightening discussions.

%%% Local Variables: 
%%% mode: latex
%%% TeX-master: "art"
%%% End: 

%\vspace{2cm}
%\pagebreak
\section*{Figures}
\begin{figure}[hhh]

     \vspace{0.5cm}
  \begin{center}
     \epsfig{file=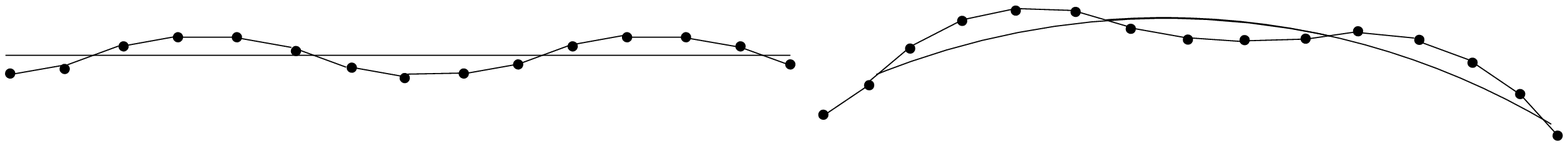,width=14cm,angle=0}
     \end{center}

      \caption{Wavy polymer with
        fixed ripple curvature amplitude on flat and bent background,
        respectively. Note the enhanced normal
        displacement from the background on the right--hand side.}
      
\end{figure}

\begin{figure}[hhh]

     \vspace{2.0cm}
  \begin{center}
     \epsfig{file=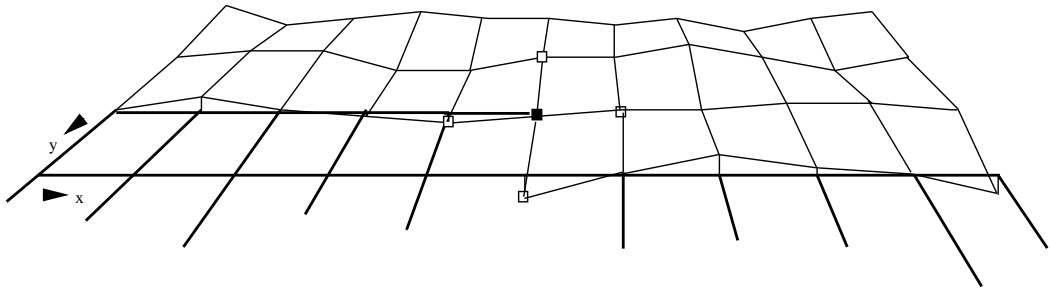,width=14cm,angle=0}
     \end{center}

      \caption{Lattice model of a membrane fluctuating around a flat base. The
        heavy lines are grid lines and shown only up to the membrane edge. The
        thin lines indicate membrane shape or, at membrane edge, normal
        displacement from the base. The five square-shaped points determine
        the mean curvature in the middle (solid square) of the plaquette}
      
\end{figure}

%%% Local Variables: 
%%% mode: latex
%%% TeX-master: "art"
%%% End: 

%\end{spacing}
%\begin{appendix}
%\input{app1}
%\end{appendix}
\vspace{10mm}
%\renewcommand{\baselinestretch}{1}

%%% Local Variables: 
%%% mode: latex
%%% TeX-master: "art"
%%% End: 


\begin{thebibliography}{99}
%\addcontentsline{toc}{chapter}{\protect\numberline{}Literaturverzeichnis}
\bibitem{Bro} Brochard F., Lennon J.-F., J. Phys. France 36 (1975), 1035
\bibitem{DGT} De Gennes P.G., Taupin C., J. Phys. Chem. 86 (1982), 2294
\bibitem{hel2} Helfrich W., J. Phys. France 46 (1985), 1263
\bibitem{hel3} Helfrich W., J. Phys. France 47 (1986), 321
\bibitem{hel4} Helfrich W., J. Phys. France 48 (1987) 285
\bibitem{P1} Peliti L., Leibler S., Phys. Rev. Lett. 54 (1985), 1690
\bibitem{For1} F\"orster D., Phys. Lett. 114A (1986), 115
\bibitem{K2} Kleinert H., Phys. Lett. 114A (1986), 263
\bibitem{ami} Ami S., Kleinert H. {\sl Functional Measures of Membrane
    Fluctuations}, Berlin preprint (1986) and\\
Ami S., Kleinert H., Phys. Lett. A 120 (1987), 207 

\bibitem{Dav1} David F. in {\sl Proceedings of the Fifth Jerusalem Winter
    School for Theoretical Physics} (1987), Nelson D.R.,Piran T. and Weinberg S.
  Eds., World Scientific, Singapore (1989)
\bibitem{Dav2} David F.,Leibler S., J. Phys. II France 1 (1994), 959 
%\bibitem{CL1} Cai W., Lubensky T.C., Nelson P. und Powers T. {\sl Measure Factors, Tension and Correlation o%f Fluid Membranes} J. Phys. II France 4 (1994), 931
\bibitem{BSK} Borelli M.E.S., Kleinert H., Schakel A.M.J. {\sl Derivative Expansion of One-Loop Effective Energy of Stiff Membranes with Tension} Berlin, preprint (1998) 




\bibitem{hel6} Helfrich W., Eur. Phys. J. B1 (1998), 481-489 
\bibitem{hel5} Helfrich W., Kozlov M.M., J. Phys. II France 4 (1994), 1427
\bibitem{GK} Gompper G. and Kroll D.M., J. Phys. I France 6 (1996), 1305
\bibitem{28c} Helfrich W., Z. Naturforschung 28 c (1973), 693
\bibitem{freed} Freed K.F. in {\sl Advances in Chemical Physics XXII}, edited
  by I.Prigogine, S.A. Rice, Wiley--Interscience (1972)
\bibitem{K3} Kleinert H., Phys. Lett. 116A (1986), 57 
\bibitem{Ou} Ou- Yang Zhong- can, Helfrich W., Phys. Rev. A Vol. 39, No. 10 (1989), 839
\bibitem{Diff3} Visconti A. {\sl Introductory Differential Geometry for Physicists}, World Scientific (1989)
\bibitem{Diff4} See textbooks on differential geometry.
\bibitem{K4} Kleinert H. {\sl Gauge Fields in Condensed Matter}, World
  Scientific (1989)

\bibitem{F} Fulling S.A. {\sl Aspects of Quantum Field Theory in Curved
    Space-Time}, Cambridge University Press (1989)

%\bibitem{hel3} Helfrich W. {\sl Size Distribution of Vesicles: The Role of the Effektive Rigidity}, J. Phys.% France 47 (1986), 321



%



%\bibitem{KR} Kegel W. K., Reiss H. {\sl Theory of Vesicles and Droplet Type Mikroemulsions: Configurational %Entropy, Size Distributions, and Measurable Properties} Berichte der Bunsengesellschaft Vol. 100, 3/1996










\end{thebibliography}
\end{document}